\documentclass[11pt,a4paper]{article}
\usepackage{jcappub}

\usepackage{graphicx}
\usepackage{amsmath}

\usepackage{txfonts}
\newcommand{\ud}{\mathrm{d}}

\notoc

\begin{document}

   \title{Estimating the redshift error in supernova data analysis}

   \author[a]{Jeong Hwa Kim}

 \affiliation[a]{Seoul Science High School\\, Jongro-gu, Seoul 110-530, Republic of Korea}
   
 \emailAdd{yhs07128@naver.com}
   
 \abstract{}{}{}{}{} 
 
  \abstract
    {Recent works have shown that small shifts in redshift -- 
   gravitational redshift or systematic errors --
   could potentially cause a significant bias in the estimation of cosmological parameters.
   I aim to verify whether a theoretical correction on redshift is sufficient to ease the tension 
   between the estimates of cosmological parameters from SNe 1a dataset and Planck 2015 results.
   A free parameter for redshift shift($\Delta z$) is implemented in the Maximum Likelihood
   Estimator. Redshift error was estimated from the Joint Light-curve Analysis(JLA) dataset and
   results from the Planck 2015 survey.
    The estimation from JLA dataset alone gives a best fit value of $\Omega_m = 0.272$,
   $\Omega_{\Lambda} = 0.390$, and $\Delta z = 3.77 \times 10^{-4}$. The best fit values of both
   $\Omega_m$ and $\Omega_{\Lambda}$ disagrees heavily with results from other observations.
   Information criteria and observed density contrasts suggest that the current data from SNe 1a is not
   accurate enough to give a proper estimate of $\Delta z$. A joint analysis with Planck results seems to give a
   more plausible value of the redshift error, and can potentially be used as a probe to 
   measure our local gravitational environment.
   }
 

   \maketitle
   \flushbottom
%

\section{Introduction}

  Observations on Type Ia Supernovae(SNE 1a) formed the
  basis for standard modeling in cosmology. Early studies\cite{riess1998observational} \cite{perlmutter1999measurements}
  conducted in the late 1990's raised the possibility of the expansion
  rate of the universe to be accelerating, which gave rise to the concept of
  `dark energy'. Since then, rapid developments were made in both
  measurements and theoretical analysis to improve the accuracy
  of parameter estimation. Empirical correlations were found among
  peak luminosity and characteristics such as light curve width\cite{phillips1993absolute}, color\cite{tripp1998two},
  and host galaxy mass(Kelly et al. 2010).
  A more strict form of statistical analysis based on bayesian hierarchical model\cite{march2011improved}
  and maximum likelihood estimator(Nielsen et al, 2016) replaced the widely-used $ \chi^2$ minimization.

  Gravitational redshift from our local environment is a convincing candidate for shifting the observed redshift.
  They can't be cancelled out unlike the effects from our peculiar velocity or gravitational redshift
  from host galaxies. But such effects were considered negligible, until
  recent works\cite{wojtak2015local} \cite{yu2013impact} showed that small errors($\Delta z \sim 10^{-5}$) in redshift measurements
  can bias the estimations of cosmological parameters by a few percent in the 
  Lambda Cold Dark Matter($\Lambda \textrm{CDM}$) model.
  Furthermore, these errors mimic some characteristics of dark energy
  and mislead observers to choose incorrect cosmological models.

  Recent estimates on cosmological parameters retrieved from SNE 1a observations 
  ($\Omega_m = 0.341, \\ \Omega_{\Lambda}
  = 0.569, H_0 = 73.03$) \cite{nielsen2016marginal} \cite{riess20162}
   disagree with Planck 2015 results ($\Omega_m = 0.309, \Omega_{\Lambda}
  = 0.691, H_0 = 67.74$) \cite{ade2016planck}, which is considered the most
  accurate observation on cosmology to date. 
  \cite{calcino2017need} attempted to alleviate this discord by adding a variable for
  systematic error in simple $\chi^2$ minimization. They came to a rather ambiguous conclusion that
  the obtained value of $\Delta z$ may not be credible due to its extremeness.
  In this paper, I attempt to clarify whether a theoretical correction is sufficient to reduce redshift errors in
  SNe 1a data analysis.

  \section{Models}
  
  \subsection{Supernova cosmology}
  
  Without correction, the distance modulus obtained from
  SNE 1a has a scatter roughly equivalent to $\pm1$ in magnitude
  and has to be corrected before being used as a distance probe.
   A widely used approach of
  `Spectral Adaptive Lightcurve Template 2'(SALT2)\cite{guy2007salt2}, a one-step
  correction method using light curve shape and color, will be used throughout this paper.
  The corrected distance modulus is:
  
 \begin{equation}
     \mu_{SN} = m^*_B - M +{\alpha}x_1 - {\beta}c
   \end{equation}
   
  Where $m^*_B, x_1, c$ each being the maximum apparent magnitude in rest frame
  B-band, the light curve width, and color. The absolute magnitude
  $M$ and constants $\alpha, \beta$ are assumed to be same for all SNe 1a.  The SN Ia distance modulus
  is then compared to the expectation in the standard $\Lambda \textrm{CDM}$ cosmological model:
  
 \begin{equation}
     \mu \equiv 25 + 5 \log_{10} (d_L/Mpc)
  \end{equation}
  
 \begin{equation}
     d_L = (1 + z)\frac{d_H}{\sqrt{\Omega_k}}\sinh\left(\sqrt{\Omega_k}
     \int_0^z{\frac{H_0{\ud z'}}{H(z')}}\right)
  \end{equation}
  
  \begin{equation}
  d_H = c/H_0, \qquad H_0 \equiv 100h \,\textrm{km}\,\textrm{s}^{-1}\,\textrm{Mpc}^{-1}
   \end{equation}
   
  \begin{equation}
     H = H_0\sqrt{{\Omega_m}(1 + z)^3 + {\Omega_k}(1 + z)^2 + \Omega_\Lambda}
  \end{equation}
  
  where $d_L, d_H, H$ are the luminosity distance, Hubble distance and Hubble parameter, and $\Omega_m, \Omega_{\Lambda}, \Omega_k$ are the matter, cosmological constant and curvature density in units of the critical density.
  The hubble parameter $h$ has a fiducial value of 0.7, but will be estimated along with other parameters.
  

\subsection{Gravitational redshift and systematic errors}

The gravitational redshift results from a difference in a gravitational potential between the point of light
reception and emission. In the weak field limit, the gravitational redshift $z_g$ is given by:

\begin{equation}
  z_g = \frac{\phi_r - \phi_e}{c^2}
     \end{equation}

$\phi_r$ and $\phi_e$ being the gravitational potential at the points of light reception and emission, respectively. 
The gravitational redshift may take a positive or negative sign, where a negative sign indicates a blueshift. 
The density contrast of the universe and the surrounding void($\delta_R$) can then be written as\cite{calcino2017need}:

\begin{equation}
\delta_R = \frac{-2 z_g c^2}{\Omega_m 100^2 h^2 R^2}
\end{equation}

A photon undergoing both cosmological and gravitational redshift has a combined redshift of:

\begin{equation}
\label{zg}
(1 + z_{tot}) = (1 + \bar{z})(1 + z_g)
\end{equation}

where $z_{tot}$ is the total redshift and $\bar{z}$ is the cosmological redshift.
This relation can be rewritten as:

\begin{equation}
\bar{z} = \frac{z_{tot} - z_g}{1+z_g}
\end{equation}

Systematic errors occur in the form of $z_{tot}=\bar{z}+\Delta z$, so small $z_g$ and $\Delta z$
have similar effects on shifting the observed redshift.

\section{Data and method}

The data used in this paper are the Joint Light-curve Analysis (JLA) sample\cite{betoule2014improved},
which contains 740 spectroscopically confirmed SN Ia in a redshift range of $0.01 < z < 1.3$
 from the Sloan Digital Sky Survey II (SDSS-II), the Supernova Legacy 
Survey (SNLS), the Hubble Space Telescope (HST), and low-redshift samples.

\subsection{Maximum likelihood estimators}

The method used in this paper is adopted from \cite{nielsen2016marginal}.
The Maximum Likelihood Estimator(MLE) is defined as:
\[\mathcal{L} = \textrm{probability density}(\textrm{Data} | \textrm{Model})\]

and for a given SN 1a, the MLE can be written as:

\begin{align}
   \mathcal{L} &= p[(\hat{m}^*_B, \hat{x}_1, \hat{c}) \,|\, \theta]\nonumber \\
  &= \int{p[(\hat{m}^*_B, \hat{x}_1, \hat{c})\, |\, (m^*_B, x_1, c),\theta] \, p[M, x_1, c) \,|\, \theta] \, \ud M \ud x_1 \ud c}
\end{align}

Where $(\hat{m}^*_B,\hat{x}_1,\hat{c})$, $(M,x_1,c)$, and $\theta$
each being the observed data, true data, and the set of cosmological parameters.
the stretch $s$, color $c$ corrections, and the absolute magnitude
$M$ are all considered as random Gaussian variables without any redshift dependence. 

This equation can be written in matrix notation
and integrated analytically to obtain:

\begin{align}
\mathcal{L} = &|2\pi(\Sigma_d + A^T{\Sigma_l}A)|^{-1/2}\nonumber\\
&\times \exp[-(\hat{Z} - Y_0A)(\Sigma_d + A^T{\Sigma_l}A)^{-1}(\hat{Z} - Y_0A)^T/2]
\end{align}

where:

\begin{displaymath}
\, \, \, A =
\left( \begin{array}{cccc}
1 & 0& 0 & \dots \\
-\alpha & 1& 0 & \dots \\
\beta & 0 & 1 &\dots \\
\vdots & \vdots & \vdots & \ddots
\end{array} \right)
\end{displaymath}
\[
      \begin{array}{lp{0.8\linewidth}}
         \Sigma_d &Experimental covariance matrix\\
         \Sigma_l &diag($\sigma^2_{M_0}, \sigma^2_{x_{1, 0}}, \sigma^2_{c_0}, \dots$)\\
        Y &\{$M_0, x_{1, 0}, c_0, \dots$\}\\
        \hat{Z} &\{$\hat{m}^*_{B1} - \mu_1, \hat{x}_{11}, \hat{c}_1, \dots$\}     
       \end{array}
\]

The most likely values of $\theta = \{\Omega_m, \Omega_{\Lambda}, \alpha, x_{1,0}, \sigma_{x_{1,0}}, \beta,
c_0, \sigma_{c_0}, M_0, \sigma_{M_0}, z_g, h\}$ are the set which maximizes the value of $\mathcal{L}$.

\subsection{Information criteria}

Information criteria(IC) tests are used to check whether complicating 
a model by adding parameters is justified.
Two widely used ICs are the Bayesian Information Criterion (BIC) \cite{schwarz1978estimating}
and the Akaike Information Criterion (AIC) \cite{akaike1998information}, and are given by:

\begin{align}
AIC &= -2\log \mathcal{L}_{max}+2N_p \\
BIC &= -2\log \mathcal{L}_{max}+N_p\log(N_{data})
\end{align}

$N_p$ and $N_{data}$ each being the number of parameters and data.
Model with a lower AIC or BIC score is considered to be better for explaining a given dataset.
They encapsulate Occam's razor by penalizing models with more parameters.
A decrease in AIC or BIC of 2 is considered positive evidence for the model with lower IC while a difference of 6 is considered strong evidence.


\section{Results and discussion}

The JLA dataset was analyzed with and without $\Delta z$, and the results are shown in Table 1. 
Flat indicates that the constraint for flat universe($\Omega_m + \Omega_{\Lambda} = 1$) is applied.
The joint analysis of JLA and Planck results(JLA \& Planck) will be discussed later.

\begin{table} [hbt]
\caption{Best fit parameters from JLA data analysis}
\small
\label{table:1}      
\centering          
\begin{tabular}{l c  c  c  c  c}
\hline\hline       
                     
Model & $-2\log \mathcal{L}_{max}$ & $\Omega_m$ & $\Omega_{\Lambda}$ & $z_g$ & $h$\\ 
\hline                
   Planck 2015 & & 0.309 & 0.691 & & 0.677 \\   
\hline 
   JLA & -214.97 & 0.340 & 0.568 & 0 & 0.702\\
   JLA with $\Delta z$ & -217.06 & 0.272 & 0.390 & $3.77 \times 10^{-4}$& 0.726\\
\hline
   JLA flat & -214.82 & 0.376 & 0.624 & 0 & 0.709\\
   JLA flat with $\Delta z$ & -215.81 & 0.392 & 0.608 & $2.16 \times 10^{-4}$ & 0.713\\
\hline
   JLA \& Planck & -209.56 & 0.309 & 0.691 & $2.60 \times 10^{-5}$ & 0.665\\
\hline
\end{tabular}
\end{table}

The addition of $\Delta z$ causes significant changes to the 
values of both $\Omega_m$ and $\Omega_{\Lambda}$,
from $\Omega_m = 0.340$ to $\Omega_m = 0.272$ and 
$\Omega_{\Lambda} = 0.568$ to $\Omega_{\Lambda} = 0.390$.
Under flat conditions, changes are smaller, resulting
$\Omega_m = 0.376$ to $\Omega_m = 0.392$ and 
$\Omega_{\Lambda} = 0.624$ to $\Omega_{\Lambda} = 0.608$.
The best fit value of $\Delta z$ is $\Delta z = 3.77 \times 10^{-4}$ for no constraints and
$\Delta z = 2.16 \times 10^{-4}$ for flat universe.
 Nuisance parameters $\{\alpha, x_{1,0}, \sigma_{x_{1,0}}, \beta, c_0,  \sigma_{c_0}, M_0, \sigma_{M_0}\}$ 
 remains mostly unaffected.

 \subsection{The validity of $\Delta z$}

 The best fit values disagree heavily with the recent 2015 Planck results.
 The total density $\Omega_m + \Omega_{\Lambda} = 0.662$
 strays from 1 and claims a less dense, less accelerating
 universe. Same can be said under flat conditions, where the increase in $\Omega_m$
 represents a less accelerating universe.
 Unlike in \cite{calcino2017need} where the increase of $\Omega_m$ shifted the estimates \emph{towards}
 Planck results, the same trend shifts the estimates \emph{away} from target value.
 The hubble constant also deviates from Planck results by increasing from $h = 0.702$
 to $h = 0.726$. The direction of this shift seems to have little connection with the actual values of
 cosmological parameters.

\begin{table} [hbt]
\small
\caption{Comparison of AIC and BIC with no $\Delta z$ as the reference}
\label{table:2}      
\centering          
\begin{tabular}{ l  c  c  c  c  c}
\hline\hline       
                     
Model & $AIC$ & $BIC$ & $\Delta (AIC)$ & $\Delta (BIC)$\\ 
\hline                    
   JLA & -192.97 & -142.30 & 0 & 0\\
   JLA with $\Delta z$ & 193.06 & -137.78 & -0.06 & 4.52\\
   \hline
   JLA flat & -192.82 & -142.15 & 0 & 0\\
   JLA flat with $\Delta z$ & -191.80 & -136.52 & 1.02 & 5.63\\
\hline
\end{tabular}
\end{table}

Table 2 shows the changes in information criteria when adding a redshift shift variable to the model.
AIC remains mostly unchanged, but increase in BIC shows that the addition of $\Delta z$
does not improve the model. This does not however, indicate that $\Delta z = 0$.
It means that shifting the redshift data does not lower the value of $\mathcal{L}$
enough to compensate the increase of model complication and suggests that the current redshift dataset
is not accurate enough to give a proper estimate of redshift error.

When $\Delta z$ is interpreted as a gravitational redshift,
the possible value of $\delta_R$ can be obtained by eq.7.
The maximum value of R should be the distance to the closest SN in the JLA dataset
having $z$ of $0.01-0.02$. Using the distance approximation at low $z$:

\begin{equation}
D = \frac{cz}{100h}
\end{equation}

gives distance of $40-80\textrm{Mpc}$. The density contrast calculated from best fit values is
then in the order of $\delta_R \sim -27.7$ to $\delta_R \sim -6.9$.
Considering the $\delta$ observed in the CMB super-void is in the scale of 
$\delta \sim 0.14$, $\delta_R$ is far too extreme to expect from our
local gravitational environment.

Overall, attempting to reduce redshift errors by directly adding a variable in the model seems
inappropriate with the current SNe 1a dataset. 

\subsection{The estimation of $\Delta z$ with Planck 2015 results}

A more accurate estimate of $\Delta z$ could possibly be achieved by
jointly analyzing with other independent measurements. 
In this section, I will present the estimation of redshift error
derived from Planck 2015 results.

The Maximum Likelihood Estimator used in the previous section can also be used
to estimate parameters under certain conditions. For example, the flat condition can be
applied by maintaining $\Omega_m + \Omega_{\Lambda} = 1$ for every iteration while
minimizing the MLE. The condition here will be fixing $\Omega_m$ and $\Omega_{\Lambda}$ to those of
Planck; $\Omega_m=0.309$ and $\Omega_{\Lambda}=0.691$.
The results are shown in Table 1. The new-found hubble constant is $h=0.665$, and $\Delta z$
is reduced to 7\% of its initial value. Estimating $\Omega_m$ and $\Omega_{\Lambda}$ again with the retrieved $\Delta z$
causes a minor shift in best fit value($\Omega_m=0.345$, $\Omega_{\Lambda}=0.580$), but is unlikely to have
much significance.

\section{Conclusion}

In this paper, I looked for evidence whether adding a free parameter for redshift error in MLE
can reduce possible bias in retrieving cosmological parameters. The estimate of $\Delta z$
from JLA dataset is $\Delta z = 3.77 \times 10^{-4}$, which corresponds to a local density contrast
almost 100 times greater than those found in CMB super-voids when interpreted as a gravitational redshift.
Increase in BIC and the deviations from Planck results suggests that the best fit values of $\Delta z$ added MLE
is not a valid estimate for the parameters $\Omega_m$, $\Omega_{\Lambda}$, and $\Delta z$.
Incorporating Planck results to the model gives a more reasonable value of $\Delta z \sim 10^{-5}$.
Jointly analyzing the SNe 1a dataset with data from other observations might reveal the matter-energy distribution
of the universe, but an independent measurement of our local environment is required to confirm its results.
This is due to the fact that the source of our redshift error is unknown; systematic errors 
and gravitational redshift -- or some other cause of error -- have very similar effects in shifting the observed redshift.

\paragraph{Code Availability}

The code and data used in the analysis are available at https://zenodo.org/record/1\\041027\#.WfvCN7b7J-U

\nocite{*}
\bibliographystyle{JHEP}
\bibliography{References}

\end{document}